\documentclass[draft,jgrga]{agu2001}
\usepackage{lineno}

\usepackage{graphicx}
\setkeys{Gin}{draft=false}

\authorrunninghead{ASAI ET AL.}
\titlerunninghead{Anemone Structure and Flares/CMEs}

\authoraddr{A. Asai,
Nobeyama Solar Radio Observatory, National Astronomical Observatory of
Japan, Minamimaki, Minamisaku, Nagano, 384-1305, Japan.
(asai@nro.nao.ac.jp)}

\linenumbers*[1]
\begin{document}

\title{Evolution of Anemone AR NOAA 10798 and the Related
Geo-Effective Flares and CMEs}

\authors{
Ayumi Asai, \altaffilmark{1,2,3}
Kazunari Shibata, \altaffilmark{4}
Takako T. Ishii, \altaffilmark{4}
Mitsuo Oka, \altaffilmark{4,5}
Ryuho Kataoka, \altaffilmark{6,7}
Ken'ichi Fujiki, \altaffilmark{7}
and 
Nat Gopalswamy\altaffilmark{8}}

\altaffiltext{1}{
Nobeyama Solar Radio Observatory, National Astronomical Observatory of
Japan, Minamimaki, Minamisaku, Nagano, 384-1305, Japan.}

\altaffiltext{2}{
National Astronomical Observatory of Japan, Minamimaki, Osawa, Mitaka,
Tokyo, 181-8588, Japan.}

\altaffiltext{3}{
The Graduate University for Advanced Studies (SOKENDAI), Japan.}

\altaffiltext{4}{
Kwasan and Hida Observatories, Kyoto University, Yamashina, Kyoto,
607-8471, Japan.}

\altaffiltext{5}{
Center for Space Plasma and Aeronomic Research,
University of Alabama in Huntsville, AL 35805, USA.}

\altaffiltext{6}{
RIKEN (The Institute of Physics and Chemical Research),
2-1 Hirosawa, Wako, Saitama, 351-0198 Japan.}

\altaffiltext{7}{
Solar-Terrestrial Environment Laboratory Nagoya University, Chikusa,
Nagoya, Aichi, 464-8601, Japan.}

\altaffiltext{8}{
NASA Goddard Space Flight Center, Greenbelt, MD 20771, USA.}

\begin{abstract}
We present a detailed examination of the features of the Active Region
(AR) NOAA 10798.
This AR generated coronal mass ejections (CMEs) that caused a large
geomagnetic storm on 24 August 2005 with the minimum Dst index of
$-216$~nT.
We examined the evolution of the AR and the features on/near the solar
surface and in the interplanetary space.
The AR emerged in the middle of a small coronal hole, and formed a {\it
sea anemone} like configuration.
H$\alpha$ filaments were formed in the AR, which have southward axial
field.
Three M-class flares were generated, and the first two that occurred on
22 August 2005 were followed by Halo-type CMEs.
The speeds of the CMEs were fast, and recorded about 1200 and
2400~km~s$^{-1}$, respectively.
The second CME was especially fast, and caught up and interacted with
the first (slower) CME during their travelings toward Earth.
These acted synergically to generate an interplanetary disturbance with
strong southward magnetic field of about $-50$~nT, which was followed by
the large geomagnetic storm.
{\it Accepted for publication in JGR.
Copyright (2008) American Geophysical Union. Further reproduction or
electronic distribution is not permitted.}
\end{abstract}

\begin{article}

\section{Introduction}
Space weather has attracted a lot of attention in recent times.
Space weather research involves various related fields, such as the
solar surface, solar wind, interplanetary space, geomagnetosphere,
ionosphere, and atmosphere, since a comprehensive understandings from
active phenomena on the solar surface to the propagation of the
disturbances toward Earth are crucially required for the studies.

Vast plasma ejected from the solar corona in the form of coronal mass
ejections (CMEs) leads to a geomagnetic storm, and therefore, CMEs have
been actively discussed.
Moreover, large geomagnetic storms are associated with large flares
(that are emitting strong X-ray), long duration events (LDEs), fast
CMEs, and so on (e.g., Gopalswamy, Yashiro, \& Akiyama 2007).
Flare locations are another factor for a major geomagnetic storm, since
the flare location close to the disk center indicates that the related
CME is heading towards Earth and is likely to cause a large geomagnetic
storm (Manoharan et al. 2004).
Ejections that cause strong disturbances with southward magnetic field
in the interplanetary space are also important.
Coronal holes (CHs) are, on the other hand, related with fast solar wind
because of their open magnetic field, and therefore, themselves have
been another important factor for space weather studies.
While large geomagnetic storms are caused by earth-directed CMEs (see
e.g., Gosling et al. 1990), weaker storms are associated with high-speed
streams from CHs (see e.g., Sheeley, Harvey, \& Feldman 1976).
However, storms related to high-speed streams from CHs cause larger flux
enhancement of MeV electrons of the Earth's Van Allen belt than the
CME-associated storms do on average (Kataoka \& Miyoshi 2006).
It has been, furthermore, reported that many fast Halo CMEs are
associated with CHs (Verma 1998; Liu \& Hayashi 2006).
The recent work done by Liu (2007) showed that the speeds are faster
even statistically than those of CMEs under the heliospheric current
sheet.
Therefore, in order to understand what kind of events can generate a
large geomagnetic storm, it is necessary to study active phenomena on
the solar surface and the propagation in the interplanetary space, in
relation to the surrounding magnetic structure.

\begin{figure}
\noindent\includegraphics[width=34pc]{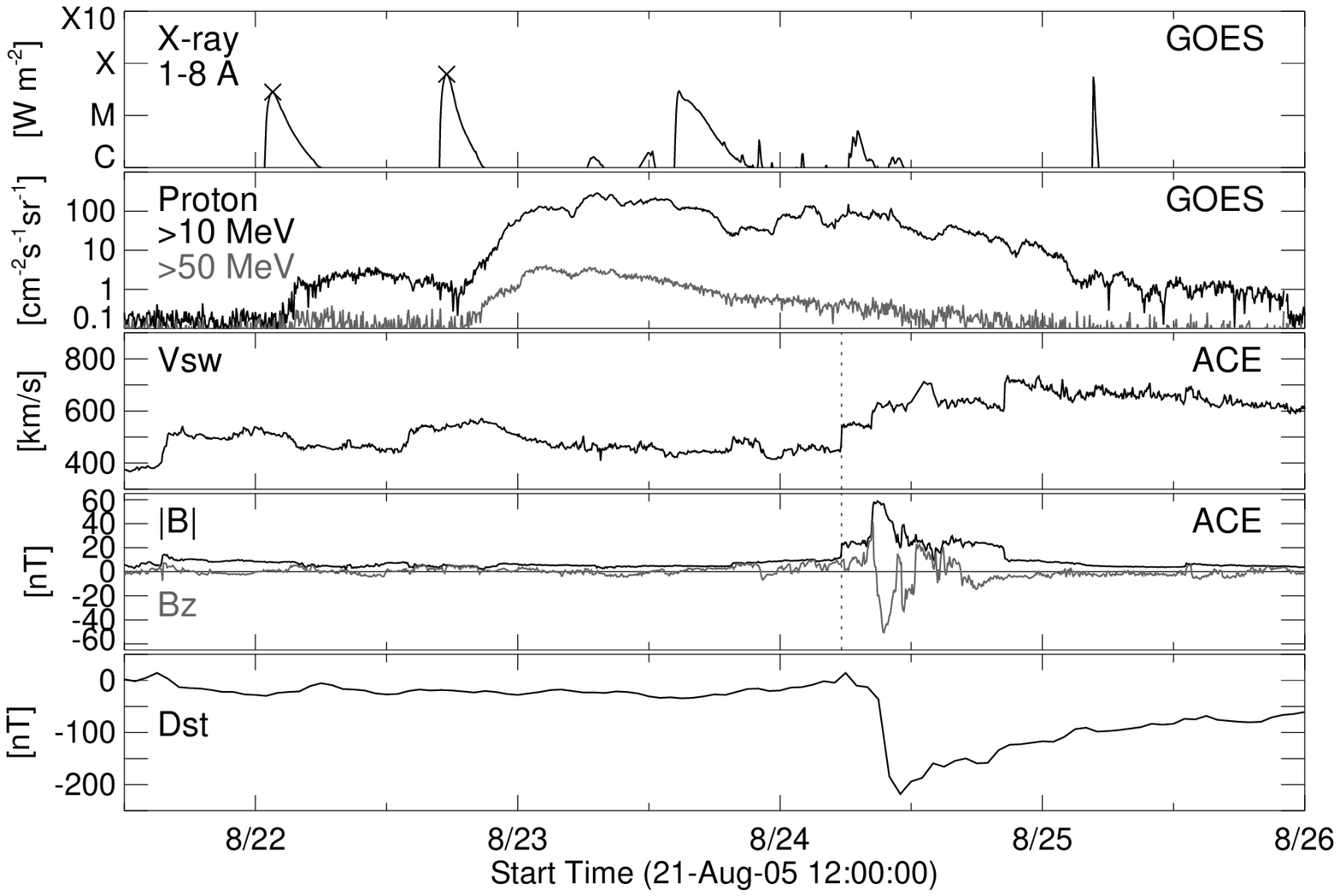}
\caption{
Overview of the geomagnetic storm that occurred on 24 August 2005, and
the related solar-terrestrial events.
{\it From top to bottom}: 
SXR flux in the {\it GOES} 1.0 - 8.0~{\AA} channel, 
proton fluxes in $>$10~MeV (black line) and $>$50~MeV (gray line)
channels obtained with {\it GOES}, 
bulk velocity of solar wind $V_{\rm sw}$ measured with {\it ACE}, 
total magnetic field strength $|B|$ (black line) and Z-component of the
magnetic field $B_{z}$ (gray line) measured with {\it ACE}, and 
Dst index produced by the Kyoto University.
\label{fig1}}
\end{figure}

In this paper we examine in detail the evolution of an Active Region
(AR) that emerged in a CH and the related flares, CMEs, and the
geomagnetic storm to elucidate how such a magnetic configuration works
to generate a geo-effective flares/CMEs.
CMEs originating from AR NOAA 10798, generated a large geomagnetic
storm on 24 August 2005, which was one of the 88 major geomagnetic
storms reported by Zhang et al. (2007).
This AR has been highly paid attention to, since it was one of the
targets of the International CAWSES\footnote{Climate And Weather of the
Sun-Earth System} Campaign\footnote{see,
http://www.bu.edu/cawses/secondcampaign.html}, and at the related 
virtual conference\footnote{http://workshops.jhuapl.edu/s1/index.html},
there were intensive discussions on the AR (e.g., Asai et al. 2006).
Figure~\ref{fig1} shows an overview of the geomagnetic storm and the
related solar-terrestrial events.
The Soft X-ray (SXR) flux in the {\it GOES} 1.0 - 8.0~{\AA} channel
shows three M-class flares that occurred on 22 and 23 August 2005.
The first two flares (marked with $\times$) were associated with the
CMEs responsible for the geomagnetic storm in question.
In the second panel we can recognize the sufficient enhancements of the
proton fluxes in $>$10~MeV (black line) and $>$50~MeV (gray line)
channels obtained with {\it GOES}.
Both flares were followed by enhancements of solar energetic particles
(SEPs), and the second flare's was larger.
The bulk velocity of solar wind $V_{\rm sw}$ in the third panel and the
total $|B|$ (black line) and Z-component of the magnetic field $B_{z}$
(gray line) in the fourth panel were measured with the {\it Advanced
Composition Explorer} ({\it ACE}).
The first shock was recorded at 05:35~UT by {\it ACE} as shown by the
dashed line.
The same shock was also recorded by the {\it Geotail} satellite at
06:15~UT.
The interplanetary magnetic field had a strong southward component of
about $-50$~nT.
The bottom panel shows the Dst index produced by the Kyoto University.
The decrease of the Dst index was quite large, reaching $-216$~nT.
In \S 2 we describe the evolution of the AR, focusing on the
photospheric magnetic configuration, and the H$\alpha$ filament formed
during the evolution of the AR and the coronal features are also
presented.
In \S 3 we discuss the flares/CMEs that occurred on 22 August 2005.
In \S 4 we shortly review the associated interplanetary disturbances.
In \S 5 we summarize our results.

\section{Evolution and Structure of Active Region NOAA 10798}
\subsection{Evolution}
\begin{figure}
\noindent\includegraphics[width=36pc]{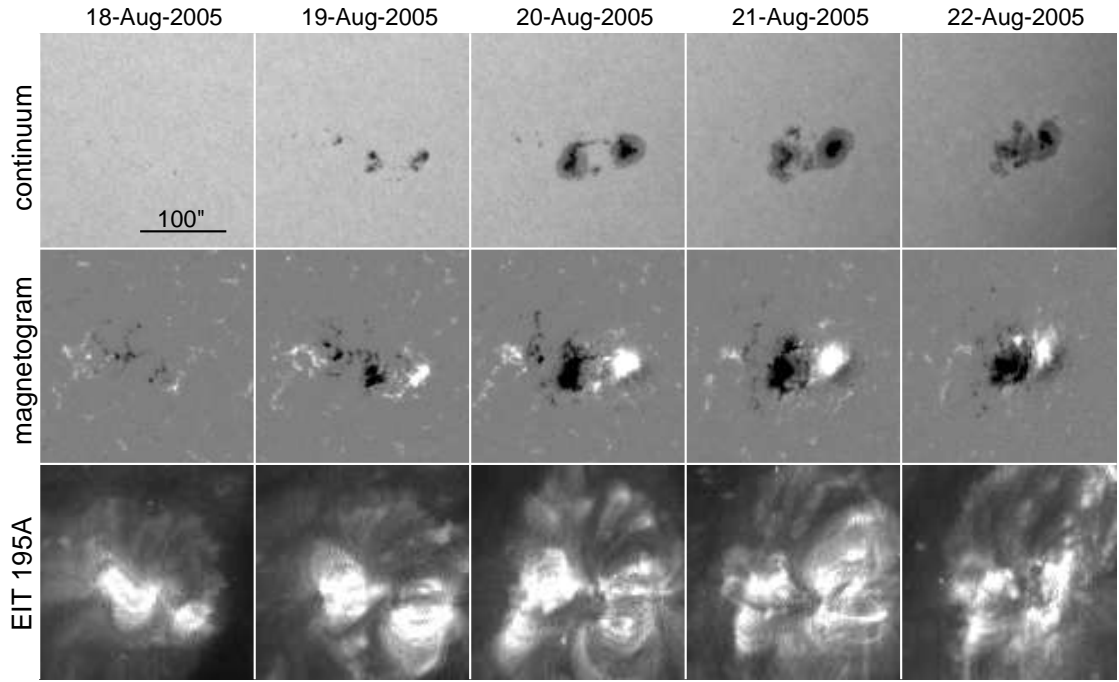}
\caption{
Temporal evolution of the AR.
The top and the middle panels show the continuum images and the
magnetograms observed with {\it SOHO}/MDI, respectively.
The bottom panel shows the EUV images obtained with {\it SOHO}/EIT.
Each image was taken at about 00:00 UT of the day.
Solar north is up, and west is to the right.
\label{fig2}}
\end{figure}

First, we examine the evolution of the photospheric magnetic field.
Figure~\ref{fig2} shows the continuum images (top panels), the
magnetograms (middle panels), and the extreme ultraviolet (EUV) images
of AR 10798.
The continuum images and the magnetograms were obtained with the
Michelson Doppler Imager (MDI; Scherrer et al. 1995) aboard the {\it
Solar and Heliospheric Observatory} ({\it SOHO}; Domingo, Fleck, and
Poland 1995), while the EUV images are taken with the
Extreme-Ultraviolet Imaging Telescope (EIT; Delaboudini\'{e}re et
al. 1995) aboard {\it SOHO}.
Each image was taken at about 00:00 UT of the day.
%
AR 10798 emerged on 18 August 2005 and rapidly evolved.

Although the region showed a simple bipolar configuration, while it was
in violation of the so-called ``Hale-Nicholson's magnetic polarity law''
(Hale et al. 1919), according to which the preceding spots in the
southern hemisphere should have a negative polarity during solar cycle
23.
These ``reverted polarity'' ARs are statistically more likely to
generate flares and CMEs (L\'{o}pez Fuentes et al. 2003; Tian et
al. 2005).
We checked all the ARs that appeared in 2005, and found that only 7
ARs (5~\%), including the AR 10798, were the ``reverted polarity'' ARs.
Furthermore, 4 of the 7 ARs, including AR 10798, showed high solar
activity.
This implies that AR 10798 had potentially a very complex structure.
For example, a highly twisted and kinked magnetic structure may be
embedded beneath the photosphere as Ishii, Kurokawa, \& Takeuchi (2000)
and Kurokawa, Wang, \& Ishii (2002) reported.
Indeed, we can see the rotating motion of the pair of the sunspots
counter-clockwise during the disk passage.
This AR further evolved and generated an X17 flare on 7 September 2005
when it returned as NOAA 10808 (Wang et al. 2006, Nagashima et
al. 2007).

\begin{figure}
\noindent\includegraphics[width=20pc]{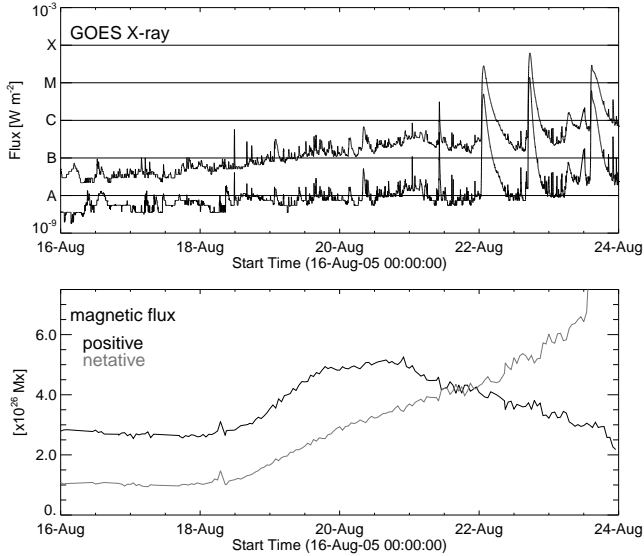}
\caption{
Time profiles of SXR flux and magnetic fluxes.
{\it Top}: SXR flux in the {\it GOES} 1.0 - 8.0 {\AA} (upper) and 0.5 -
4.0 {\AA} (lower) channels.
{\it Bottom}: Magnetic flux of the AR observed with {\it SOHO}/MDI.
The calculated area is $400^{\prime\prime} \times 400^{\prime\prime}$
centered on the middle of the AR, and is as wide as it covers the whole AR.
The time profile of the negative magnetic flux is multiplied by $-1$.
\label{fig3}}
\end{figure}

The top panel of Figure~\ref{fig3} shows the SXR lightcurves obtained by
{\it GOES} in the 1.0 -- 8.0 {\AA} (upper) and 0.5 -- 4.0 {\AA} (lower)
channels.
The bottom panel of Figure~\ref{fig3} shows the time profiles of the
magnetic flux of this AR.
That for the negative magnetic flux is multiplied by $-1$.
The calculated area is $400^{\prime\prime} \times 400^{\prime\prime}$,
and is as wide as it covers the whole AR.
Following the emergence and evolution of the active region from 18
August 2005, the magnetic fluxes as well as the SXR intensity
gradually increased, with three M-class flares occurring on 22 and 23
August, before rotating behind the west limb.
In this paper we mainly discuss the first two flare that occurred on 22
August 2005, since the geomagnetic storm on 24 August 2005 is attributed
to the related eruptions/CMEs.

\subsection{Filament Formation}
Second, we examine the filament formation in AR 10798, using the
H$\alpha$ images.
A filament is just a visualized part of a helical flux rope, and it is
only a fraction of the whole of the flux rope.
However, it is thought that it is located in the middle of the
flux rope, and that a filament even represents the whole structure
(see e.g., Low \& Hundhausen 1995).
Therefore, an eruption of a filament is always related with a large
scale disturbance of the coronal magnetic field that often appears as
an EIT dimming.
As Munro et al. (1979) suggested that more than 70\% of CMEs are
associated with eruptive prominences or filament disappearances (with or
without flares), and therefore, filament eruptions are very important as
a CME-associated phenomenon.
Moreover, we often see an ejected filament observed in
H$\alpha$s/microwaves or a plasmoid in SXRs at the center of an ejecta,
when it is accompanied by a flare.
We can even roughly extrapolate the magnetic configuration of a CME
from that of the ejected filament.
For example, Rust (1994) showed that the helicity of ejected filaments
correspond to the chirality of magnetic clouds passing Earth (about 4
days after the eruptions).

\begin{figure}
\noindent\includegraphics[width=32pc]{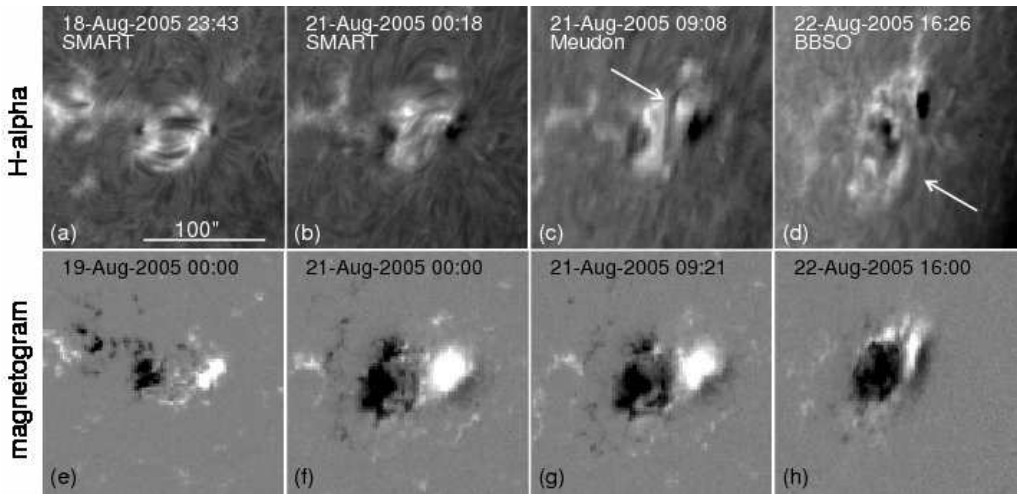}
\caption{
{\it Top}: H$\alpha$ images.
(a) and (b) were obtained with SMART at Hida Observatory, Kyoto
University.
(c) and (d) were obtained at Observatoire de Paris, Section de Meudon
and Big Bear Solar Observatory, respectively.
{\it Bottom}: Magnetograms taken with {\it SOHO}/MDI.
\label{fig4}}
\end{figure}

Figure~\ref{fig4} shows the temporal evolution of the AR in the
H$\alpha$ images (top panels) and the magnetic field (bottom panels).
The H$\alpha$ images in Figure~\ref{fig4}a and~\ref{fig4}b were taken
with the Solar Magnetic Activity Research Telescope
(SMART\footnote{http://www.hida.kyoto-u.ac.jp/SMART/}) at Hida
Observatory, Kyoto University.
Figure~\ref{fig4}c and~\ref{fig4}d were obtained at the Observatoire de
Paris, Section de Meudon and the Big Bear Solar Observatory,
respectively.
Both of these images were obtained through the on-line data center of
the Global High-Resolution H$\alpha$
Network\footnote{http://www.bbso.njit.edu/Research/Halpha/}.
The magnetograms (Fig.~\ref{fig4}e -~\ref{fig4}h) were taken by {\it
SOHO}/MDI.

For three days after the emergence of the AR (until 20 August), a clear
arch-filament system (Bruzek 1967) was seen (Fig.~\ref{fig4}a).
The bipole-like systems bridged the neutral line and connected the spots
of opposite polarity.
On the other hand, after 21 August 2005, these filamentary structure was
abruptly changed.
In Figure~\ref{fig4}b, some oblique structure appeared, and showed
pre-filamental structure.
Comparing with the magnetograms (the bottom panels of
Figure~\ref{fig4}), we confirm that the magnetic field of these structure
was oriented from northwest to southeast, which means they had southward
magnetic field.
About nine hours after this (Fig.~\ref{fig4}c), the sheared filamentary
structure evolved to a large H$\alpha$ filament that lay on the magnetic
neutral line between the sunspots.
The arrow in Figure~\ref{fig4}c points to the filament.
The filament formation is consistent with what Martin (1973) pointed out
long ago: developed filaments usually become apparent about the ``fourth
day'' after the initial formation of an active region.
Unfortunately, there are no data between Figure~\ref{fig4}b to
\ref{fig4}c, but we can see the new flux emergence around the magnetic
neutral line, (compare Figure~\ref{fig4}f with~\ref{fig4}g).
The formation of the filaments with the southward magnetic field is 
probably related to the emerging flux.

The first M-class flare occurred at 00:44 UT on 22 August 2005, which is
in the middle of the time between Figure~\ref{fig4}c and~\ref{fig4}d.
In Figure~\ref{fig4}d, we can recognize the disappearance of the
H$\alpha$ filament after the first flare, while a new filament formed in
the south part of the AR as pointed by the arrow in the panel.
Associated with the first flare, the H$\alpha$ filament formed in the
northern part (Fig.~\ref{fig4}c) erupted, and with the second flare, the
southern one (Fig.~\ref{fig4}d) erupted.
The axial field of these filaments had southward magnetic field, which
is easily inferred from the pre-filamentary structure.
Although we also checked the EUV data taken with EIT and the
Transition Region and Coronal Explorer, we could not find out any
phenomena that can be a source of the CMEs other than the filament
eruptions.
We will discuss the flares and CMEs in more detail in \S 3.

\subsection{Anemone Structure}
\begin{figure}
\noindent\includegraphics[width=28pc]{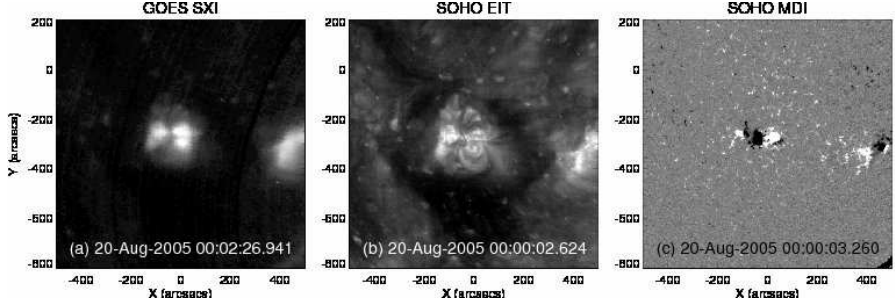}
\caption{
Coronal feature of AR 10798.
(a): A SXR image obtained with {\it GOES}/SXI.
(b): An EUV (195{\AA}) image obtained with {\it SOHO}/EIT.
The bright region near the center of the image is the AR.
The surrounding dark region is a CH.
(c): A magnetogram taken with {\it SOHO}/MDI.
\label{fig5}}
\end{figure}

Figures~\ref{fig5}a,b show the coronal structure of the AR observed at
about 00:00~UT on 20 August 2005, in SXR and in EUV with Solar X-ray
Imager (SXI) on board {\it GOES} and {\it SOHO}/EIT, respectively.
The bright structure near the center of the image is AR 10798.
Figure~\ref{fig5}c shows the magnetogram taken by {\it SOHO}/MDI.
The following sunspot with the negative (that is {\it black}) polarity
is the center of the EUV bright structure, and a radial array of loops
is formed.
We also present schematic cartoons of the magnetic structure of AR 10798
in Figure~\ref{fig6}a and b.

The appearance is like a {\it sea anemone}, and this type of ARs is
sometimes called ``anemone structure'' (Shibata et al. 1994a, 1994b), or
originally ``fountain'' (Tousey et al. 1973, Sheeley et al. 1975a) in
the {\it Skylab} era.
We call these ARs ``anemone ARs'' in this paper.
These anemone ARs are often associated with the emerging fluxes within
unipolar regions (Sheeley et al. 1975b), and in most cases, they appear
in CHs (Asai et al. 2008).
Although characteristics of anemone ARs have been mainly discussed
only in SXRs, they are commonly seen under such a magnetic
configuration, even in a chromospheric line by the Solar Optical
Telescope on board {\it Hinode} (Shibata et al. 2007).

As shown in Figure~\ref{fig5}, AR NOAA 10798 is clearly surrounded
with a unipolar region with the positive magnetic polarity, and shows
the anemone appearance both in SXR and in EUV, and we can conclude
that NOAA 10798 was a typical anemone AR.
In emerging, the AR interacted (reconnected) with the ambient coronal
field, and magnetic loops were arranged radially with the following spot
that has the negative magnetic polarity as the center of the anemone
structure.
In Figure~\ref{fig5}b the dark region surrounding the AR is a CH.
On 22 August, when the flares/CMEs in the matter occurred, the anemone
appearance somewhat changes as seen in Figure~\ref{fig2}.
This is caused by projection like many anemone ARs, while some anemone
ARs keep the appearance even on the limb (Saito et al. 2000).

\section{Flares and CMEs}
Next, we focus on the two M-class flares and the associated CMEs.
The first flare that was M2.6 on the {\it GOES} scale, started at 00:44
UT, and peaked at 01:33 UT.
The second one was M5.6 on the {\it GOES} scale, and the start and the
peak times were 16:46 UT and 17:27 UT, respectively.
Both are LDEs, and showed clear arcade structure.
Figure~\ref{fig6}c shows a schematic of the magnetic field during the
flares.
The sites of the flares were (S11$^{\circ}$ W54$^{\circ}$) and
(S12$^{\circ}$ W60$^{\circ}$), respectively.

The two flares were associated with disappearances of the H$\alpha$
filaments, and Halo-type CMEs that were observed with the Large Angle
Spectrometric Coronagraph (LASCO) aboard {\it SOHO} (see the {\it
SOHO}/LASCO CME online
catalog\footnote{http://cdaw.gsfc.nasa.gov/CME\_list/}; Yashiro et
al. 2004).
LASCO images of the two CMEs (CME1 and CME2) are shown in
Figure~\ref{fig7}.
The left panels are the LASCO C2 running difference images overlaid with
EUV images taken by {\it SOHO}/EIT (195{\AA}), and the right panels are
the LASCO C3 running difference images.
CME1 was ejected mainly to the northwest, and CME2 was to the southwest
as indicated by the arrows in Figure~\ref{fig7}.
The directions were roughly consistent with the initial position of the
H$\alpha$ filaments (see, \S 2.2).

It is particularly notable that the CMEs were quite fast: CME1 and CME2
had speeds of about 1200 and 2400~km~s$^{-1}$, respectively.
The speed of CME2 is ranked among the top 17 of all the 13,000 CMEs
observed by {\it SOHO}/LASCO until the end of 2007.
Although the time interval between the two flares/CMEs was about 16
hours, CME2 possibly caught up with CME1 before reaching 1~AU
(Gopalswamy et al. 2001a).
Statistically, a CME ejected with the velocity of $V_{\rm CME}$ have an
acceleration $a$ m~s$^{-2}$ $= 2.193 - 0.0054 \times V_{\rm CME}$
km~s$^{-1}$ (Gopalswamy et al. 2001b), and the expected accelerations
for CME1 and CME2 are $-4.3$ and $-10.8$~m~s$^{-1}$ for the current
case.
Therefore, we estimate that the interacting between CME1 and CME2
occurred at about 1~AU (i.e. near Earth), by assuming constant
accelerations for the CMEs.

The interplanetary disturbance associated with the Halo-type CMEs can be
followed by using interplanetary scintillation (IPS).
When we see a radio source through a highly turbulent plasma associated
with a CME traveling from the Sun, the radio source scintillates.
Therefore, such scintillations show us the the electron density
fluctuation caused by the CME.
As an effective indicator of the electron density fluctuation, we often
use g-value ($g$) calculated from IPS data (see Tokumaru et al. 2000,
2003, 2005 for more details).
The g-value represents the variation of the electron density fluctuation
in the solar wind ($\Delta{N_e}$), as  
$g^2\propto\int_0^{\infty}\Delta{N_e^2}w(z)dz$, 
where $z$ is a distance along the line-of-sight, and $w(z)$ is the IPS
weighting function given by Young (1971). 
It is normalized to the mean level of density fluctuations so that the
quiet solar wind yields g-value around unity, and the enhancement ($g >
1$) shows the passing of a turbulent plasma.

We examined the g-values taken with IPS at Solar-Terrestrial Environment
Laboratory (STEL), Nagoya University (Kojima \& Kakinuma 1990, Asai et
al. 1995, Tokumaru et al. 2000).
Figure~\ref{fig8} shows the daily (Japanese daytime) sky projection maps
of the g-values.
In each map, the center corresponds to the location of the Sun, and
dotted cocentric circles are constant radii contours from the Sun drawn
at 0.3, 0.6, and 0.9 AU.
The solid circles indicate the points of the closest approach to the Sun
(P-points) on the line-of-sight where g-value were obtained (P-point
approximation).
The locations of the stronger g-values are emphasized by colors and
sizes of the circles.
The dark gray and the black circles represent the locations where the
g-values are larger than 1.5 and 2.0, respectively.
In the both panels of Figure~\ref{fig8} we can see density fluctuations
that were caused by the two Halo CMEs, while we cannot distinguish the
individual CMEs discretely due the low spatial resolution of IPS.
The front of the disturbance, which was caused by CME1, reached about
0.4 and 0.8~AU on August 23 and 24, respectively.
CME1 is well decelerated, and the speed is about 700~km~s$^{-1}$.
These fluctuations are distributed roughly in all direction.

\section{Interplanetary Disturbance}
Here, we investigate in more detail why a strong disturbance with a
magnetic field of about $-50$~nT arrived at Earth.
As we mentioned above (see \S 4), CME2 probably caught up with CME1, and
therefore, disturbance is regarded as a merged product of CME1 and CME2,
although the interaction was not directly observed.

Figure~\ref{fig9} shows a 7-hour interval corresponding to the
geo-effective part of the interplanetary disturbance from {\it
Geotail}.
The top four panels present the magnetic field in GSE coordinates
obtained by the magnetic-field experiment (MGF; Kokubun et al. 1994).
The magnitude $|B|$ and the x- ($B_x$), y- ($B_y$), and z-components
($B_z$) of the magnetic field are shown.
The fifth panel shows the ion velocity $V_x$ observed with the low
energy particle experiment (LEP; Mukai et al. 1994).
The bottom panel shows the electron density N$_e$ observed by the plasma
wave instrument (PWI; Matsumoto et al. 1994).
The density reached so high that the counts of the particle detectors
onboard {\it Geotail} (and probably {\it ACE} as well) were saturated,
and therefore, it is underestimated during the storong disturabance.
To avoid the underestimation of the density,
we simply traced local enhancements of the electrostatic noise that
appears in the dynamic spectra of the electric field as have been
carried out elsewhere (see e.g., Fig. 4 of Terasawa et al., 2005).
Although the measurement also have a uncertainty, it is more accurate
than that by particle measurement experiments, since the counts were
not saturated.

A flux rope (FR) structure can be identified by the smooth rotation of
the magnetic field from 09:15 to 11:15 UT as shown by the vertical
dashed lines in Figure~\ref{fig9}.
It is notable that the 2 hours duration of this FR was extremely short
compared to the typical duration of about 20 hours (Lepping et al. 2003;
Gopalswamy 2006).
The local velocities of the FR is 650~km~s$^{-1}$, as will be discussed
below.
We estimated the radial size of the FR to be about 0.03~AU, by
multiplying the local velocity by the 2 hours duration.
This value is also extremely small compared with typical ones of 0.2 -
0.3~AU (Forsyth et al. 2006).
This FR showed the smooth rotation from positive to negative $B_{y}$, a
negative $B_{z}$ peak in the middle of the $B_{y}$ rotation, and
relatively small $B_{x}$ component.
These can be roughly explained by the passage of a right-handed flux
rope with the southward pointing axis field, which is consistent with
the magnetic field configuration of the associated H$\alpha$ filaments.
The largest geomagnetic storm of cycle 23 that occurred on 20 November
2003 was associated with a similar FR (Gopalswamy et al. 2005).

About 15 minutes before the front edge of the FR (i.e. at about 09:00
UT), a solar wind discontinuity is identified by the sudden increases in
the magnetic field, solar wind speed, and density as shown by the
vertical solid line.
From the variation of velocity distribution function, we confirmed an 
abrupt increase of temperature (not shown) there, and concluded that the
discontinuity is a shock.
We call the discontinuity as the ``second shock'', and the ``first
shock'' is for the one observed at the beginning of the event as shown
with the dashed line in Figure~\ref{fig1} and the vertical solid line at
about 06:15~UT in Figure~\ref{fig9}.

The extremely strong southward magnetic field, the unusual short
duration of the FR (2 hours), and very small separation between the
second shock and the FR front (15 minutes) can be naturally explained,
if we regard the disturbance as the product of very fast shock wave
associated with CME2 interacting with the slower body of CME1 in
traveling to Earth.
Therefore, CME2 suffered from a strong deceleration, which implies that
there was a great compression of the interplanetary medium in front of
CME2.
In this case the first and the second shocks are thought to be
associated with CME1 and CME2, respectively.

The local velocities of the first and the second shocks are measured
by the positional relation between the {\it ACE} and {\it Geotail}
satellites, and therefore, we can roughly estimate their accelerations
(decelerations).
{\it ACE} and {\it Geotail} were located at (223.7, 10.6, 4.5) and
(12.9, 25.7, 1.9) R$_E$ ($= 6378$ km) at 09:00~UT on 24 August 2005 in
GSE coordinate system, respectively.
As already mentioned, CME1 and CME2 were ejected with velocities of
about 1200 and 2400~km~s$^{-1}$, at intervals of 16 hours.
On the other hand, the local velocities of the first and the second
shocks are 650 and 710~km~s$^{-1}$, respectively, and the time
separation between them is reduced to only about 3 hours.
Assuming the constant accelerations, they are estimated to be $-3.2$
and $-13$~m~s$^{-2}$.
As we calculated above, the accelerations for CME1 and CME2 are
statistically expected to be $-4.3$ and $-10.8$~m~s$^{-1}$.
The additional deceleration of CME2 also indicates that it interacted
with slower CME1 and compressed the interplanetary medium there.

\section{Summary and Discussions}
In order to make clear the importance of an AR that emerged in a CH to
generate geo-effective flares/CMEs, we examined the evolution of the
AR NOAA 10798, the solar events associated with a geomagnetic storm
that occurred on 24 August 2005, and the related interplanetary
disturbances.
The summary of the features of the AR and the events is as follows:
(1) Highly twisted and complex magnetic flux emerged within a small CH
on 18 August 2005, which was named NOAA 10798,
(2) An anemone type structure was generated, and H$\alpha$ filaments
that had southward axial fields were formed on 21 August 2005, 
(3) Two halo CMEs associated with M-class flares occurred on 22 August
2005.
(4) The CME speeds were fast, especially the second one recorded 2400
km~s$^{-1}$,
(5) The interplanetary disturbances with strong southward magnetic field
of about $-50$~nT and strong compression of plasma were produced.

The CMEs were particularly geo-effective, and the minimum Dst index was
$-216$ nT.
The reasons for the CMEs to be so geo-effective were the high speeds of
the two CMEs and their interaction as well as the CMEs traveled
directly toward the earth.
For the current case, the speed of CME2 was faster and pushed the slower
CME1, which led to a unusual strong compression of the plasma at the
front of CME2.

The high speeds of the CMEs are more notable.
The AR was large and very complex, and violated the Hale-Nicholson's
magnetic polarity law.
These reverted polarity ARs are statistically favorable to produce
large flares.
However, it is suspicious whether just the violation of the
Hale-Nicholson's law is responsible for high speed CMEs of about
2000~km s$^{-1}$, and it should be quantitatively and statistically
clarified in the future.
In this paper we suggest that the fast CMEs are probably a consequence
of the eruption inside a CH from an anemone AR.
This is consistent with the association between fast Halo CMEs and CHs
as reported before (Verma 1998; Liu \& Hayashi 2006; Liu 2007).

Eruptive activities of anemone ARs are usually low (Asai et al. 2008),
and often confined to small-scale activities inside CHs that appears to
be SXR bright points.
In some cases, anemone ARs can produce large SXR coronal jets (Shibata
et al. 1994b, Vourlidas et al. 1996, Kundu et al. 1999, Alexander \&
Fletcher 1999).
This is because the situation of an emergence of a magnetic flux within
a CH is suitable for magnetic reconnection with the surrounding field to
generate SXR coronal jets and/or H$\alpha$ surges (Yokoyama \& Shibata
1995, 1996).
Wang (1998) indicates the possibility that even polar plumes are
associated with jets from anemone ARs at high latitudes.
Anemone ARs are related to non-radial coronal streamers emanating from
magnetically high latitudes (Saito et al. 2000).
The relation between anemone ARs and fast solar winds have also been
paid attentions to (Takahashi et al. 1994, Saito et al. 1994, Wang 1998).
Saito et al. (2000) further discussed the rotational reversing model and
the triple dipole model to explain the reversal of the solar surface
magnetic field, and anemone ARs play an important role in this.
This model implies that anemone ARs are more conspicuous in the decaying
phase of a solar cycle as in the case of AR NOAA 10798.

\begin{figure}
\noindent\includegraphics[width=28pc]{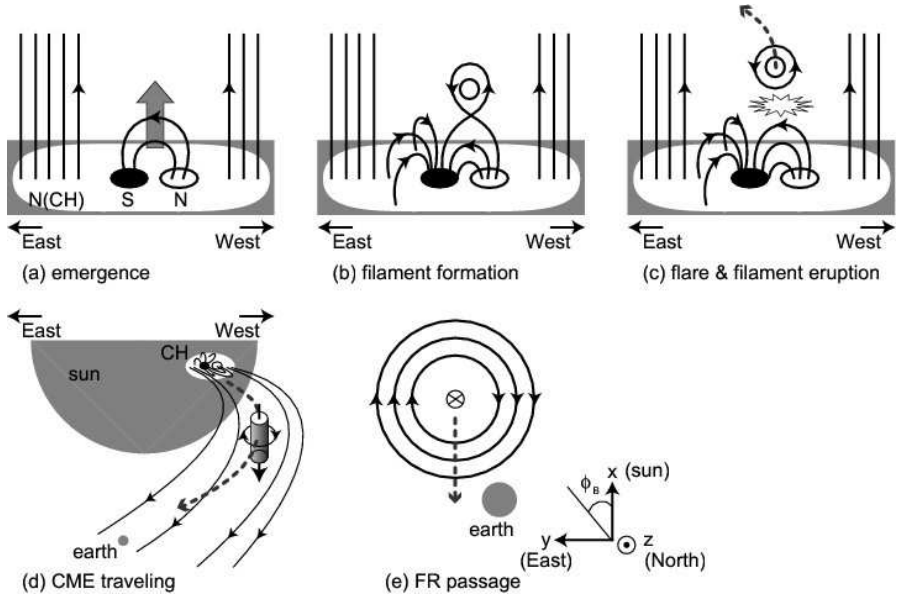}
\caption{
Schematic cartoon of AR 10798 and related flares/CMEs.
(a) A magnetic flux newly emerges within a CH.
(b) An anemone structure is generated, and an H$\alpha$ filament is also
formed above the emerged flux.
(c) A magnetic reconnection occurs beneath the filament, which causes
the filament eruption.
The ejected plasma is bent eastward by the surrounding magnetic field
with positive magnetic polarity.
(d) The ejecta becomes a magnetic cloud (shown as a cylinder) that have
a southward axial magnetic field and is approaching to Earth.
(e) Passage of a FR and the variation of the azimuthal angle of the
magnetic field $\phi_{B}$.
When a FR passes the east of the earth, $\phi_{B}$ evolutes 90 -- 180 --
270 ($-$90) degree.
\label{fig6}}
\end{figure}

\begin{figure}
\noindent\includegraphics[width=20pc]{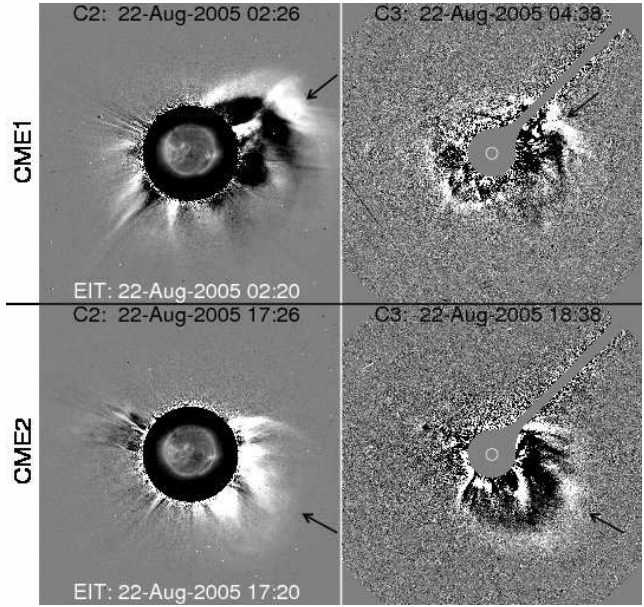}
\caption{
White-light CME images obtained with {\it SOHO}/LASCO.
The left two panels show C2 running difference images for CME1/CME2
overlaid with EUV images obtained with {\it SOHO}/EIT (195{\AA}).
The right two panels show C3 running difference images.
The arrows roughly point the main part of CMEs.
\label{fig7}}
\end{figure}

\begin{figure}
\noindent\includegraphics[width=20pc]{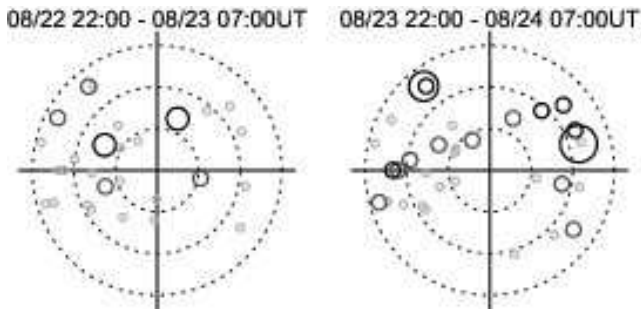}
\caption{
Daily (Japanese daytime) sky projection maps of g-values obtained with
IPS observations at STEL Nagoya University.
In each map, the center corresponds to the Sun center, and the dotted
cocentric circles are constant radii contours from the Sun drawn at
0.3, 0.6, and 0.9~AU.
Solid circles indicate the points of the closest approach to the Sun
(P-points) on the line-of-sight where g-value data were obtained
(P-point approximation).
Dark gray and black circles represent the locations where the g-values
are larger than 1.5 and 2.0, respectivelly.
\label{fig8}}
\end{figure}

\begin{figure}
\noindent\includegraphics[width=32pc]{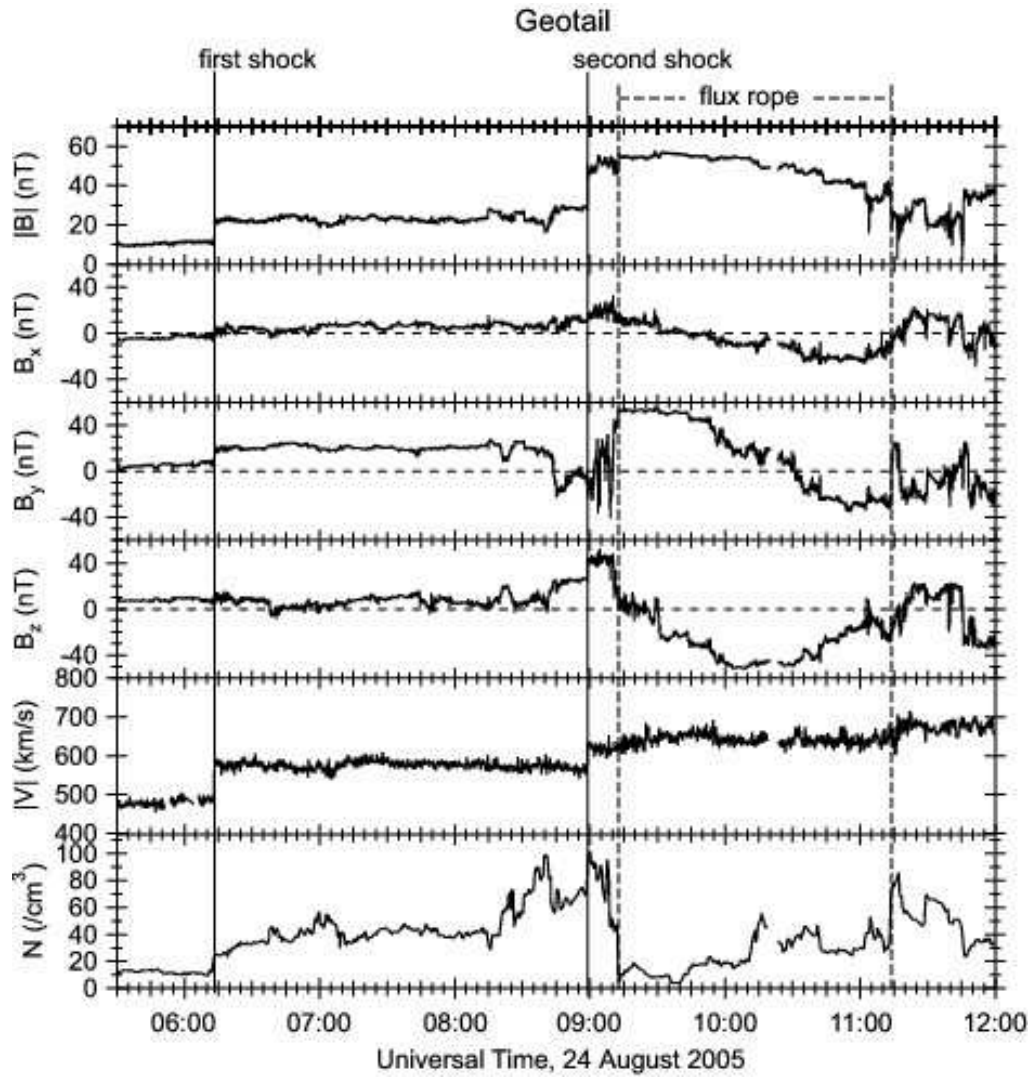}
\caption{
{\it Geotail} observation of the interplanetary disturbance on 24 August
2005.
{\it From top to bottom}: magnitude and X-, Y-, and Z- components in GSE
coordinate of the magnetic field (MGF experiment), 
ion velocity (LEP/SWI experiment), and electron density (PWI/SFA).
The vertical solid lines show the shocks (the first and the second
shocks).
The vertical dashed lines show the flux rope.
\label{fig9}}
\end{figure}

On the other hand, the deflection of CMEs eastward by the interplanetary
fields effectively worked in the current case as shown in the
Figure~\ref{fig6}d.
As Wang et al. (2004) pointed, the faster CMEs are deflected more
eastward, and therefore, the AR NOAA 10798 generated geo-effective CMEs,
although it was quite close to the southwest limb.
The azimuthal angle of the magnetic field measured from the x-axis
$\phi_{B}$ ($= \arctan(B_{y}/B_{x})$) changed 90$^{\circ}$ --
180$^{\circ}$ -- 270$^{\circ}$ ($-$90$^{\circ}$) during the passage of
the FR, which is consistent with the guess that the deflection of the
CMEs were so strong that the axis of the FR passed through the east of
the earth.
This is also consistent with the fact that the flares in the next
rotation (and renamed as NOAA 10808) did not affect the magnetosphere so
much (Wang et al. 2006, Nagashima et al. 2007).
Furthermore, the extremely short duration of FR and the missing of CME1
(see Fig.~\ref{fig9}) are possibly explained by the skimming encounter
with the CMEs due to the strong deflection.

The nature of the interplanetary disturbances and their impact on the
magnetosphere strongly depend on the features of emergence and evolution
of an AR and the relation with the surrounding magnetic field.
In this work we succeeded to follow in detail the evolution of the AR and
the large geomagnetic storm resulting from eruptions in the AR.
The reconstruction of the proposed scenario using numerical simulations
will be attempted in the future.


\begin{acknowledgments}
This work was supported by the Grant-in-Aid for the Global COE Program
``The Next Generation of Physics, Spun from Universality and Emergence''
from the Ministry of Education, Culture, Sports, Science and Technology
(MEXT) of Japan.
This work was also supported by the Grant-in-Aid for Creative Scientific
Research ``The Basic Study of Space Weather Prediction'' (17GS0208, Head
Investigator: K. Shibata) from the Ministry of Education, Science,
Sports, Technology, and Culture of Japan.
This work was partially carried out by the joint research program of the
Solar-Terrestrial Environment Laboratory, Nagoya University.
We would like to acknowletge all the members of the {\it Geotail}/PWI,
LEP, and MGF for providing the data.
We would like to thank WDC for Geomagnetism, Kyoto Dst index service.
Our thanks also go to the SMART teams of Hida Observatory, Kyoto
University, Big Bear Solar Observatory, and Meudon Observatoire de
Paris, Section de Meudon for letting us use the H$\alpha$ data.
We made extensive use of {\it SOHO}, and {\it ACE} Data Center.
MO was supported by the Grant-in-Aid for JSPS Postdoctoral Fellows for
Research Abroad.
\end{acknowledgments}


\end{article}


\end{document}